\documentclass[aps,groupedaddress,superscriptaddress,amsmath,amssymb,twocolumn,prb]{revtex4-2}
\usepackage{bm}
\usepackage[pdftex]{graphicx}
\usepackage{xcolor}
\usepackage{color}
\usepackage[normalem]{ulem}
\usepackage{enumerate}
\usepackage{epstopdf}
\usepackage{braket}
\usepackage{hyperref}
\usepackage{caption}
\usepackage{blindtext}

\begin{document}

\title{Spin-dependent recombination mechanisms for quintet bi-excitons generated through singlet fission}

\author{Yan Sun}
\affiliation{LPS, Universit\'e Paris-Saclay, CNRS, UMR 8502, F-91405 Orsay, France}
\author{L. R. Weiss}
\affiliation{Pritzker School of Molecular Engineering, University of Chicago, Chicago, IL, USA}
\author{V. Derkach}
\affiliation{O. Ya. Usikov Institute for Radiophysics and Electronics of NAS of Ukraine 12, Acad. Proskury st., Kharkov, 61085, Ukraine }
\affiliation{LPS, Universit\'e Paris-Saclay, CNRS, UMR 8502, F-91405 Orsay, France}
\author{J. E. Anthony}
\affiliation{Department of Chemistry, University of Kentucky, Lexington, KY 40506-0055, USA }
\author{M. Monteverde}
\affiliation{LPS, Universit\'e Paris-Saclay, CNRS, UMR 8502, F-91405 Orsay, France}
\author{A.D. Chepelianskii}
\affiliation{LPS, Universit\'e Paris-Saclay, CNRS, UMR 8502, F-91405 Orsay, France}


\begin{abstract}
We investigate the physical mechanisms for spin-dependent recombination of a strongly bound pair of triplet excitons generated by singlet fission and forming a spin quintet (total spin of two) bi-exciton. For triplet excitons the spin-dependent recombination pathways can involve intersystem crossing or triplet-triplet annihilation back to the singlet ground state. However the modeling of spin-dependent recombination for quintets is still an open question. Here we introduce two theoretical models and compare their predictions with the broadband optically detected magnetic resonance spectrum of a long lived quintet bi-exciton with known molecular structure. This spectrum measures the change in the fluorescence signal induced by microwave excitation of each of the ten possible spin transitions within the quintet manifold as function of a magnetic field. While most of the experimental features can be reproduced for both models, the behavior of some of the transitions is only consistent with the quintet spin-recombination model inspired by triplet intersystem crossing which can reproduce accurately the experimental two-dimensional spectrum with a small number of kinetic parameters. Thus quantitative analysis of the broadband optically detected magnetic resonance signal enables quantitative understanding of the dominant spin-recombination processes and estimation of the out-of equilibrium spin populations.
\end{abstract}


\maketitle

Singlet fission is a carrier multiplication process specific to some
organic materials in which a singlet exciton can decay into two lower energy triplet excitons
each with an energy around half of the singlet state. Recently this process has shown promise to
increase the efficiency of solar cells \cite{Hanna2006,tayebjee2012,Ehrler2012,lee2013,Rao2017} allowing them to reach efficiencies beyond the Shockley-Queisser limit \cite{Congreve2013,Dirk2018,Einzinger2019}. Recent progress and promising pathways for fission-enhanced photovoltaics are reviewed in \cite{Tayebjee2022,Hudson2022}.  It is believed that during singlet fission a transient, bi-exciton is formed mediating the transition between the photoexcited singlet and two dissociated triplet excited states \cite{Smith2010,Swenberg1968,kazzaz1968}. A similar transient triplet pair is also believed to act as an intermediate state during the reverse process of triplet-triplet annihilation. This reverse process leads to upconversion where two low-energy triplet excitons merge into one higher energy singlet excitation  \cite{Keivanidis2003,Singh2010,Schmidt2010,Pandey2015,Wenping2021,Bossanyi2021}, with potential applications in catalysis \cite{Ravetz2019,Castellano2011} and bio-imaging \cite{Liu2012,Liu2018}.

While bi-exciton excited states play a key role in a range of materials where quantum size effects are important \cite{Bryant1990,PhysRevLett.64.1805,PhysRevLett.88.117901,Baldo2000,Klimov1998,Masui2013}, their optical characterization is challenging. Their spectroscopic properties are difficult to predict theoretically, in organic materials their spectra typically overlap with singly-excited states making it hard to discriminate using conventional
spectroscopic techniques. Observation of intermediate states was for example possible using time-resolved photo-emission spectroscopy \cite{Chan2011} and optical pump-probe spectroscopy which suggested ultra-fast separation \cite{Wilson2011} with possible multi-exciton transient intermediates \cite{Pensack2016,Bakulin2016,Monahan2017}. Spin resonance techniques can resolve higher spin states formed through spin-spin coupling between electron, holes or triplet-excitons \cite{teki2000intramolecular,teki2001pi,teki2008design,bayliss2015spin}. Light-induced spin resonance thus played a key role to demonstrate that stable bi-excitons can be formed from singlet fission providing evidence of higher spin S=2 states \cite{Tayebjee2017,Weiss2017}. These states are called quintets in the spin resonance literature. The strong exchange interaction between the two bound triplet excitons was then confirmed by high magnetic fields experiments showing level anti-crossings between singlet S=0, triplet S=1 and quintet S=2 manifolds of the triplet pair \cite{bayliss2018,bayliss2016spin,wakasa2015can,chakraborty2014massive,huang2021competition,yago2022triplet}. 
Recently S=2 bi-exciton states were observed in several systems, including several reports investigating quintet pairs at room temperature \cite{Lubert-Perquel2018,nagashima2018singlet,matsui2019exergonic,chen2019quintet,bae2020spin,XU2022528} and their quantum properties can be investigated in detail with double resonance magnetic resonance techniques \cite{bayliss2020probing}.

An insight into the molecular structure of the bi-exciton state is provided through the spin properties of the quintet state
which can be used as a ruler to identify the possible pair configurations in a given system.
Indeed the inter-spin distance and orientation of the triplet change its spin dipole-dipole interaction providing a
pathway to determine the geometry of spin-pairs, by analogy with spin-label technique in in biological and synthetic systems \cite{di2014porphyrin}.
Conveniently, the dipole-dipole interactions between the two triplet states in the quintet Hamiltonian provides a native probe of its
spatial confinement and orientation \cite{weil2007electron,benk1981theory}. Note that due to the transient nature of the quintets their structural properties cannot be studied more directly through conventional X-ray techniques, and so far spin resonance has provided the most direct characterization of their microscopic structure. Since the spin dipole interaction is a small correction to the total quintet fine structure,
distinguishing it from other contributions can be difficult in conventional fixed frequency magnetic resonance techniques.
Optical detection of magnetic resonance is very sensitive \cite{shinar2012optically,clarke1982triplet,schmidt1968optical} down to single molecule level \cite{wrachtrup1993optical,kohler1993magnetic} and does not require an electro-magnetic cavity to enhance the sensitivity of microwave detection, this technique can thus be implemented in a broadband frequency range 
allowing to measure directly the dependence of the spin energy spectrum on the magnetic field. We applied this technique in TIPS-tetracene, a reference system that exhibits highly efficient singlet fission \cite{stern2015identification,stern2017vibronically}. This allowed us to show that in this system the two bound triplet excitons reside on nearby $\pi$-stacked TIPS-tetracene dimers, the good agreement between structural parameters obtained from spin resonance and crystallographic data indicated that the stability of these bi-excitons was not defect mediated and characteristic of the pristine system. More recently this approach was also used by another group to study bi-exciton structure a newly synthesized fission material \cite{Gajadhar}.

A seeming paradox underlies the observation of quintet states in optically detected magnetic resonance: why are these 'dark' states visible? Contrast in ODMR requires a spin-dependent emission mechanism -- a connection between the spin states in the "dark" quintet manifold and the observed photoluminescence intensity.
The strong exchange interaction within the quintet bi-exciton restricts direct radiative recombination via the singlet state because a change of the total spin from S=2 to S=0 would be required for triplet-triplet annihilation to occur.
Yet, some spin-dependent relaxation pathways remain for the quintet despite their strong exchange energy, providing the contrast in optical life-times for quintet spin eigen-states  which is detected in the optically detected magnetic resonance (ODMR) experiment. The microscopic mechanisms behind this spin selectivity are not understood currently without first principle theoretical understanding or accurate phenomenological models. Here we compare two possible spin-dependent recombination mechanisms which can provide ODMR contrast for the quintet bi-exciton. One such pathway, relies on matrix elements only inside the quintet manifold and can be viewed as an extension of triplet intersystem crossing models \cite{kraffert2014charge,budil1991chlorophyll,el1972pmdr,Singh1965}, the second considers the mixing with the singlet state triplet pair configuration which can be induced by spin-Hamiltonian fine-structure terms. 
We attempt for discrimination between these two possible mechanisms by modelling the ODMR signal amplitudes theoretically and comparing with experiment. While the ODMR amplitude of most of the transitions can be reproduced accurately within both models, some transitions are particularly sensitive to the underlying mechanism allowing to discriminate between the two rate models. We discuss the physical origin of this difference in the predicted signal amplitudes for the two models. We conclude by showing that once the spin-dependent recombination model is fixed broadband optically detected magnetic resonance can allow us to reconstruct the occupation probabilities of the spin states inside the quintet, which may have applications in finding systems exhibiting population inversion for maser applications.

\section{Broadband ODMR spectroscopy}

\begin{figure}[h]
\centerline{
\includegraphics[clip=true,width=9cm]{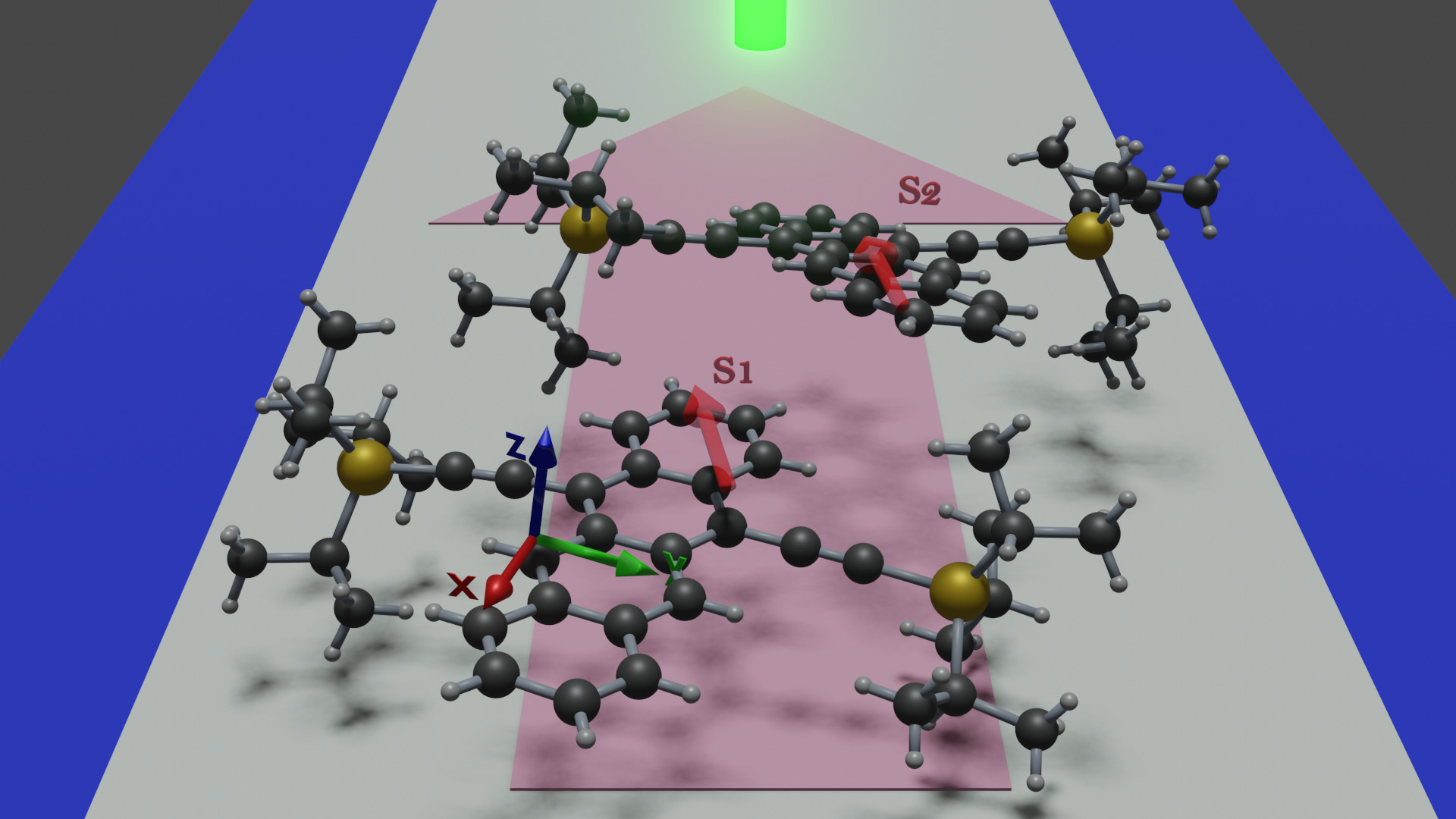} 
}
\caption{Schematic view of the experimental setup. A TIPS-tetracene crystal is placed on a microwave stripe-line microwave (white) with the pink arrow showing the direction of the AC current creating the microwave magnetic field for spin-resonance. The pi-stacked dimer of TIPS-tetracene, show the molecular configuration of a pair of triplets excitons forming an effective spin-quintet (S=2) biexciton. This exciton pair is induced through singlet fission due to illumination from ~532 nm laser through an optical fiber right next to the crystal, the same fiber is also used to collect the photoluminescence which is used for optically detected magnetic resonance experiments. The two aligned triplet spins $S_1$ and $S_2$ are shown as transparent red arrows, and the molecular basis for one of the TIPS-tetracene molecules is shown as the $x,y,z$ basis. } 
\label{odmrSketch}
\end{figure}

To analyze the spin-dependent recombination rates we continued experiments on the model quintet compound TIPS-tetracene \cite{bayliss2014geminate,ODMR2019} for which the quintet effective spin Hamiltonian model is highly accurate, as confirmed once more by the new experiments shown below. This material choice allows us to focus on the problem of modeling of the ODMR signal amplitudes for a known spin Hamiltonian. 
For the broadband optically detected magnetic resonance (ODMR) experiment a TIPS-tetracene crystal (bis(triisopropylsilylethynyl)-tetracene) was deposited on a microwave stripline, which provided the AC magnetic field for magnetic resonance excitation. A sketch of the setup is shown on Fig.~\ref{odmrSketch}. Optical excitation was performed by using 532 nm laser through an optical fiber which was also used for the fluorescence readout. Changes in the fluorescence were then detected with a lock-in amplifier at a typical microwave square-wave modulation frequency of 137Hz. To achieve a higher sensitivity in the two dimensional microwave frequency/magnetic field map compared to our previous experiments \cite{ODMR2019}, we used a side-group deuterated sample providing narrower magnetic resonance linewidths and thermalised the sample near 4K in Helium gas instead of immersing it into liquid helium which induces a strong bubbling noise in the optical readout if the helium is not superfluid. The good crystalline quality of the sample is confirmed by the sharp ODMR transition lines which do not broaden substantially with magnetic field for most transitions. The direction of the magnetic field coincides with a high accuracy with the $c$ crystal direction (\cite{ODMR2019}).

\begin{figure*}
  \centering
  \begin{minipage}{0.45\textwidth}
        \centering
        \includegraphics[clip=true,width=9cm]{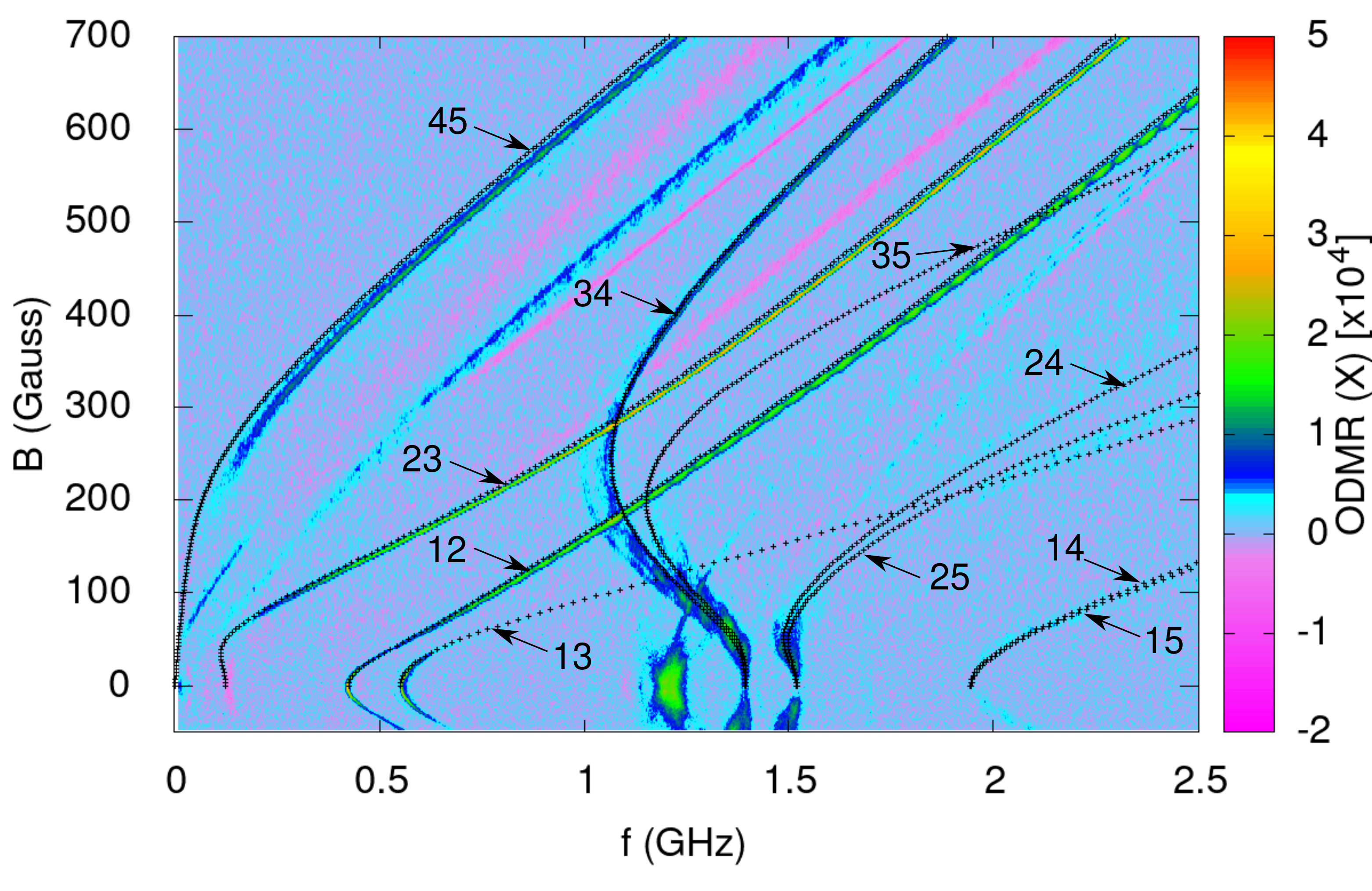}         
        \caption{Two dimensional optically detected magnetic resonance (ODMR) spectrum showing the ODMR spectrum (color scale) as function of magnetic field $B$ and microwave excitation frequency. Black lines shows the position of the expected transitions for the quintet $\pi$ stacked bi-exciton model (a sketch of this configuration is shown on Fig.~\ref{odmrSketch}, and the energy spectrum of this model is displayed on Fig.~\ref{odmrLevels}).}
        \label{odmrExp}
  \end{minipage}\hfill
  \begin{minipage}{0.45\textwidth}
        \centering
        \includegraphics[clip=true,width=8.5cm]{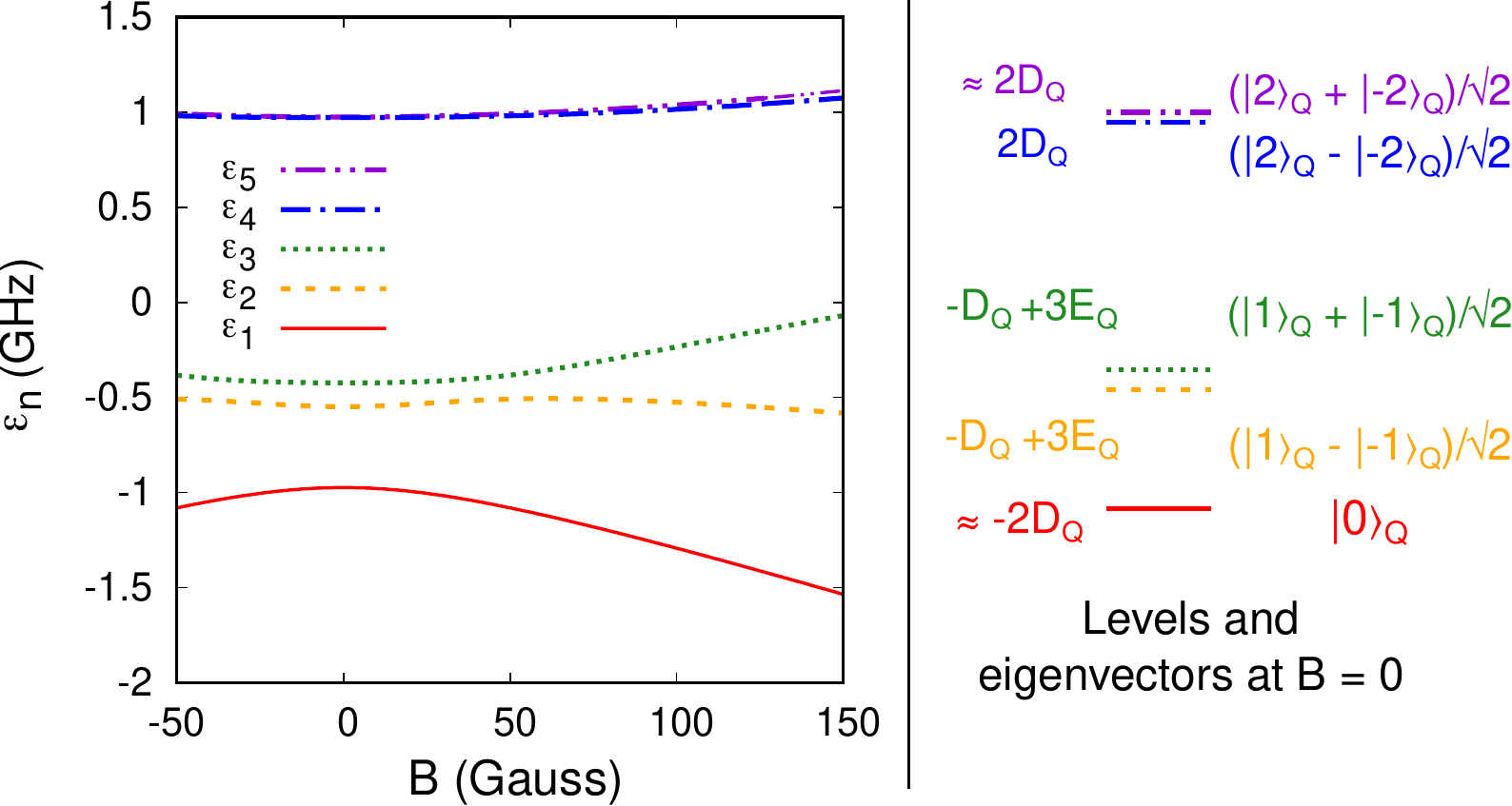} 
        \caption{Evolution of the theoretical eigenspectrum as function of the magnetic field for the $\pi$-stacked bi-exciton quintet identified in Fig.~\ref{odmrExp}. Right hand-side shows the theoretical eigenvalues and eigenvectors in the $D_Q \gg E_Q$ approximation.}
        \label{odmrLevels}
  \end{minipage}
\end{figure*}

The magnetic resonance spectrum of the quintet bi-exciton can be described from the effective zero field spin-Hamiltonian
\begin{align}
  {\hat H}_Q = D_Q {\hat S}_z^2 + E_Q ({\hat S}_x^2 - {\hat S}_y^2)
  \label{eq:H0}
\end{align}  
the energy spectrum at finite magnetic field is obtained by adding a Zeeman term to the zero field Hamiltonian ${\hat H}_Q$. The excellent agreement between this spin Hamiltonian and the experimental spin transitions is seen on Fig.~\ref{odmrExp}. On this figure the ODMR amplitude signal is shown as a color map as function of magnetic field and microwave excitation frequency. Dark lines highlight the expected transitions for the spin Hamiltonian Eq.~\ref{eq:H0} for a microscopic quintet configuration with a bi-exciton pair on nearby $\pi$-stacked TIPS-tetracene molecules with an excellent agreement with the position of the transitions observed in the experiment. In addition to explaining the fine structure parameters $D_Q = 474\;{\rm MHz}$ and $E_Q = 22.5\;{\rm MHz}$ a microscopic model based on this geometry also explains the more subtle rotation of the quintet zero field basis Eq.~\ref{eq:H0} with respect to the molecular basis of the molecules ( see Fig.\ref{odmrSketch} ). The geometric arguments underlying this rotation and the microscopic derivation of the quintet spin Hamiltonian were already presented \cite{ODMR2019} and take into account the direction between the two triplets which gives the direction of the spin-dipole coupling. Here we focus on the understanding of the ODMR signal amplitudes for a known quintet spin-Hamiltonian. Finally we note that the experimental spectrum Fig.~\ref{odmrExp} features some non-identified transitions which can be due to triplets or triplet pairs with a different geometry, we will not discuss possible assignment of these peaks in this paper. 

In Fig.~\ref{odmrExp} transitions between spin-eigenstates i and j are labelled 'ij' where the corresponding energy levels and their dependence on magnetic field are shown in Fig Fig.~\ref{odmrLevels}. The energy differences between these eigenstates are also plotted as black-lines on Fig.~\ref{odmrExp}. Fig.~\ref{odmrLevels} also shows the zero field energy spectrum and eigenvectors of the Hamiltonian Eq.~(\ref{eq:H0}) in the approximation $D_Q \gg E_Q$. The approximate expressions for the eigenvectors in Fig.~\ref{odmrLevels} are given in the zero-field quintet basis in which Eq.~\ref{eq:H0} is written, this is emphasized by the subscript $|.\rangle_Q$ in the quintet-kets. In a simplified model where the zero field basis of both triplets and quintet all coincide, the expression of quintet spin-eigenstates would be:
\begin{align}
|2\rangle_Q &= |1,1\rangle \\
|1\rangle_Q &= \frac{1}{\sqrt{2}} \left( |1,0\rangle +  |0,1\rangle \right) \\
|0\rangle_Q &= \frac{1}{\sqrt{6}} \left( |1,-1\rangle + 2 |0,0\rangle + |-1,1\rangle \right)
\end{align}
where $|s_{za}, s_{zb}\rangle$ gives the triplet state with similar expressions for $|-1\rangle_Q, |-2\rangle_Q$.

In the present case, the expressions are only approximate because of the small $30\deg$ tilt between the molecular and quintet basis. To avoid confusion, all numerical calculations are done with the exact eigenvectors of the microscopic bi-exciton model, the approximate expressions are only used for qualitative interpretation of the numerical results. For completeness we note the main ingredients of this model, it is presented in more detail in \cite{ODMR2019}. The triplet pair spin Hamiltonian reads:
\begin{align}
  \hat{H}=\sum_{\nu=a,b}{\hat S}_{\nu}^{T}\cdot\mathbf{D}_{T}^{\nu}\cdot {\hat S}_{\nu}-\Delta(\mathbf{u}_{ab}\cdot\mathbf{{\hat S}}_{a})(\mathbf{u}_{ab}\cdot\mathbf{{\hat S}}_{b})-J\mathbf{{\hat S}}_{a}\cdot\mathbf{{\hat S}}_{b}
  \label{Htt}
\end{align}
Here $\nu=a,b$ labels the two paired triplet excitons and $\mathbf{D}_{T}^{\nu}$ is their fine structure tensor. This tensor is characterized by the main fine structure parameter $D_T = 1450\;{\rm MHz}$ which is oriented along the molecular $z_\nu$ direction (see Fig.~\ref{odmrSketch}) giving the direction of $\pi$-orbitals of the corresponding TIPS-tetracene molecule. The remaining fine structure parameter $E_T$ describing the anisotropy in the molecular $x,y$ plane is very small $\le 22$MHz \cite{ODMR2019}, its effects on the quintet energy level is within experimental accuracy and we take it as zero for simplicity. The constant $\Delta = \frac{3\mu_{0}\mu_{B}^{2}g^{2}}{4\pi r_{ab}^{3}} = 51.5\;{\rm MHz}$ gives the strength of the dipole dipole interaction ($r_{ab} = 1\;{\rm nm}$ is the distance between the two TIPS-tetracene molecules) and the unit vector $\mathbf{u}_{ab}$ gives the direction. Finally in the last term, $J$ gives the amplitude of the exchange interaction. In the limit $J \gg D_Q$ the precise value of the exchange energy becomes irrelevant, high magnetic field experiments suggest that $J \simeq g \mu_B\; 30 \;{\rm Tesla}$ so this approximation is well justified. 
The distances and directions are taken from the reported TIPS-tetracene crystal structure, which for a single TIPS-tetracene dimer is shown on Fig.~\ref{odmrSketch}. Thus for a given triplet-pair geometry there are no adjustable parameters in this Hamiltonian and yet it explains the experimental 2D-ODMR spectrum very accurately. Somewhat unexpectedly, this model also allows us to fix the sign of $D_T > 0$.  Indeed the $\Delta>0$ is known from microscopic dipole-dipole interaction and thus $D_T > 0$ and $D_T < 0$ lead to different fine structure parameters $D_Q$ and $E_Q$, only positive $D_T$ is consistent with the experiment. From this argument we believe that the spin eigenstates in Fig.~\ref{odmrLevels} are indeed labeled in the order of increasing energy $\epsilon_1 < \epsilon_2 < ... < \epsilon_5$. The asignement $D_T > 0$ is consistent with results obtained using transient magnetic resonance \cite{Weiss2017}.

As shown in Fig.~\ref{odmrExp}, the experimental ODMR amplitudes are positive for all 10 possible transitions for the quintet manifold, except for transition 23 which is negative near zero magnetic field and changes to positive at higher magnetic fields. This sign change will be one of the important ingredients to discriminate between the two possible spin-dependent recombination pathways that we analyzed theoretically. The second ingredient will be the selection-rule-forbidden transitions $14$ and $15$ which correspond to direct transitions between $|0\rangle_Q$ and $|\pm2\rangle_Q$ which are expected to be strongly suppressed by spin resonance selection rules and yet are visible in the experiment.
We note that non-labelled transitions likely correspond to other triplet or multi-exciton states, but their identification and characterization is outside the scope of this paper.

\section{Kinetic models for ODMR amplitudes}

In order to interpret the amplitude of the ODMR experiment a kinetic model is needed to describe the equilibrium spin populations and how they are changed by microwave induced spin resonance transitions between the spin-eigenstates. Since we rely on optical detection of magnetic resonance, we also need to predict the contrast, ie. fluorescence yield which is expected for each eigenstate. Previously, in cases where the spin-dependent recombination pathways were identified we tried as far as possible to solve the quantum master equation as this is a more microscopic approach. Here since our purpose is instead to identify the main spin-dependent recombination pathways, we use a simplified rate equations approach which is valid when relaxation is much slower that the underlying quantum dynamics. Probably the simplest kinetic model suitable to describe our experiment will include for each spin eigen-state $|n\rangle$, a spin-dependent (radiative) recombination rate $\Gamma_n$ and a spin independent non radiative rate $\Gamma$:
\begin{align}
  \partial_t P_n = -\Gamma_n P_n - \Gamma P_n + \alpha + \alpha_n
  \label{eqrate}
\end{align}
the terms  $\alpha_n$  and $\alpha$ give the spin-dependent/independent generation rate. For simplicity we have not considered inter-conversion between spin states in this kinetic model. For the purpose of comparison of our results with pulsed optical spin resonance experiments a few words of caution. These generation rates $\alpha$ and $\alpha_n$ must include all the spin-relaxation effects which are relevant on the time scale of the spin-dependent radiative recombination which provides contrast for the ODMR experiment. The corresponding generation rates will thus be different from the generation rates which can be estimated using transient electron spin resonance a short time after pulsed light excitation. Analyzing the link between these two quantities is not a straightforward task. 

For the simple rate equation Eq.~(\ref{eqrate}), the steady state population is given by:
\begin{align}
P_n = \frac{\alpha_n + \alpha}{\Gamma_n + \Gamma}
\end{align}
with the total fluorescence yield:
\begin{align}
PL = \sum_n \frac{(\alpha_n + \alpha) \Gamma_n}{\Gamma_n + \Gamma}.
\end{align}
We assumed that all the spin-dependent recombination rates $\Gamma_n$ all correspond to radiative recombination. This is motivated by the strong singlet selection rule for optical processes in materials with weak spin-orbit interaction, while recombination through defects is in general expected to be less spin-selective. For example for trap-mediated recombination, we can expect the exciton to be trapped first regardless of its spin and then decay on a time-scale longer than spin-relaxation. 

Pumping the magnetic resonance transition between states $n,m$ tends to equilibrate the spin populations. At saturation, both $P_n$ and $P_m$ will be equal to the average population.
\begin{align}
&P_n \rightarrow (P_n+P_m)/2\\ 
&P_m \rightarrow (P_n+P_m)/2
\end{align}

We can use this to estimate the ODRM signal from transition $n \rightarrow m$, under saturated conditions:
\begin{align}
  ODMR_{nm} 
 &= \frac{(\Gamma_m - \Gamma_n)(P_n - P_m)}{2} \label{eq:ODMR}
\end{align}

To find the total ODMR signal we then sum over all possible spin transitions multiplying the amplitude $ODMR_{nm}$ by microwave absorption ${\cal A}_{nm} \ge 0$ from the transition $n \rightarrow m$.
\begin{align}
  ODMR = \sum_{n < m} ODMR_{nm} \times {\cal A}_{nm}
\label{eq:ODMR2}
\end{align}
We used the following expression for ${\cal A}_{nm}$ based on a two level approximation for the transition $n \rightarrow m$,
\begin{align}
A_{nm} = \frac{|\langle n|\Omega_{ac} \mathbf{u}_{ac} \cdot \mathbf{{\hat S}}|\rangle m|^2 }{ \gamma_0^{-2} + (\omega - |\epsilon_n - \epsilon_m|)^2 }  
\label{eq:ODMR3}
  \end{align}
where $\gamma_{0}$ is the spin lifetime and $\mathbf{u}_{ac}$ is the direction of the AC magnetic field created by the strip-line and $\Omega_{ac}$ is proportional to the AC magnetic field amplitude.
Here $\gamma_0$ is a fixed transition independent spin-lifetime. In the experiment the linewidths are influenced by spin dephasing times, inohomogenous broadening and structural inhomogenities of the sample. Thus we did not attempt to obtain a quantitative agreement between the transition linewidths in the experiment and in the simulations. Our aim was to compare the predictions of two different possible kinetic models, and the qualitative conclusions we will draw on the difference in ODMR contrast between the two models for some transitions will not be sensitive to the way the theoretical amplitude of the transitions can be rescaled by broadening effects.

To proceed further with the estimation of ODMR amplitudes we need to connect the rate equations with the spin Hamiltonian and its eigenbasis $\{|n \rangle\}$. One possibility is to assume that the spin-dependent rates are given by the overlap of the spin-eigenstates with the singlet state $|S\rangle$ of the triplet pair. In this case, we can write $\alpha_n  = \alpha_s |\langle S|n\rangle|^2$ and $\Gamma_n = \Gamma_s  |\langle S|n\rangle|^2$ where the singlet wavefunction is:
\begin{align}
|S\rangle = \frac{1}{\sqrt{3}}\left( |1,-1\rangle - |0,0\rangle + |-1,1\rangle\right)
\end{align}
If instead spin-orbit interaction mediated intersystem crossing becomes dominant, we would need instead to consider the expectation value of a projector onto the zero field basis. However as long as we consider a single spin-dependent recombination mechanism, we will have the relation:
\begin{align}
  \alpha_n &= \frac{\alpha_s}{\Gamma_s} \Gamma_n
  \label{eq:propto}
\end{align}

It turns out that relation Eq.~(\ref{eq:propto}) imposes a very strict restriction on the sign of the ODMR signal.
Injecting this equation into Eq.~(\ref{eq:ODMR}) we find: 
\begin{align}
  ODMR_{nm} &= \frac{(\alpha \Gamma_s - \alpha_s \Gamma)(\Gamma_m - \Gamma_n)^2}{2 \Gamma_s (\Gamma+\Gamma_m)(\Gamma+\Gamma_n) } 
\end{align}
This equation captures correctly the qualitative rule of a negative ODMR signal in a regime dominated by geminate pairs ($\alpha_s \Gamma > \alpha \Gamma_s$) and of a positive ODMR signal for spin-dependent recombination of non-geminate pairs ($\alpha_s = 0$ and $\Gamma_s > 0$) \cite{frankevich1982triplet,bayliss2014geminate}. However the sign of the signal depends only on the difference $\alpha \Gamma_s - \alpha_s \Gamma$. It is independent of the spin Hamiltonian and thus cannot be changed by the magnetic field, in contrast with the experiment.

It thus seems that relation Eq.~(\ref{eq:propto}) is too restrictive. A possibility to explain the sign change observed in the experiment for transition 23 is instead to consider a combination of several spin-dependent mechanisms with different weights for the emission and recombination channels. In this case the simple proportionality relation Eq.~(\ref{eq:propto}) would no longer be valid and the kinetic model becomes sufficiently general to describe a magnetic field dependent sign change in the ODMR signal. Such an assumption seems also very reasonable physically since singlet fission is a very fast phenomenon, while all spin-dependent recombination processes have a much slower kinetics. Indeed these processes need to be slow enough to allow the spin to precess under microwave pulses which is a 100-ns to $\mu$s time-scale at our microwave powers. Thus hot carrier effects and molecule structural relaxation will all be very different for generation and recombination channels.

We now propose two possible recombination models for the observed ODMR amplitudes. Focusing first on a spin-dependent intersystem crossing model, we assume recombination rates which are given by Eqs.~(\ref{eq:recS2gen}\ref{eq:recS2}). We will refer to the following as {\it Model (I) }.
\begin{align}
  \alpha_n  &= \sum_{i \in x,y,z} a_{i} \langle n | (\mathbf{u}_{Qi} \cdot \mathbf{{\hat S}}) ^2|n \rangle \label{eq:recS2gen}
\\
  \Gamma_n  &=  \sum_{i \in x,y,z} r_{i} \langle n | (\mathbf{u}_{Qi} \cdot \mathbf{{\hat S}}) ^2 |n \rangle
\label{eq:recS2}
\end{align}
where the vectors $\mathbf{u}_{Qx},\mathbf{u}_{Qy},\mathbf{u}_{Qz}$ form the quintet zero-field basis in which the zero-field Hamiltonian has the simple form given in Eq.~(\ref{eq:H0}). 
The vector $\mathbf{{\hat S}}$ is the total spin and the constants $a_{x,y,z}$ and $r_{x,y,z}$ gives the weight of each recombination channel. This model is inspired by the phosphorescence rates for triplet excitons which are selective in the zero field basis. More formally, this means that phosphorescence rates will be given by the expectation values of the projectors on the zero field basis states. Taking the example of the projector on the $|0_z\rangle$ state (the subscript $z$ emphasizes the quantization axis), we find that the projector is given by:
\begin{align}
  |0_z\rangle \langle 0_z| = 1 - {\hat S}_z^2 = \frac{S_x^2 + S_y^2 - S_z^2}{2}
  \label{eq:Pz}
\end{align}
which has the same form as equation Eqs.~(\ref{eq:recS2gen}, \ref{eq:recS2}). We then assume that for a strongly bound pair of triplet excitons forming a quintet the form of Eq.~(\ref{eq:Pz}) remains valid replacing the individual spin of the triplet exciton by the total spin of the exciton pair. This form is explicitly independent of the basis choice for the spin operators, this allows us to use the same form of the spin-dependent recombination rates for both zero and non-zero magnetic fields. Without a microscopic theory, it is also possible to express relaxation rates based on the projectors on the quintet spin basis with vanishing projection on the molecular axis. This choice would correspond to introducing forth powers of the spin operators into Eq.~(\ref{eq:propto}) and we have thus decided to limit ourselves to Eq.~(\ref{eq:propto}) for simplicity. 

We now present a second possible spin-dependent recombination model, which we refer to as {\it Model (II)}. This spin-dependent mechanism is given by the projection of spin eigenvectors onto the singlet state $|S\rangle$. For a strongly bound quintet state with a large exchange energy between the triplets the matrix elements $|\langle n|S\rangle|$ are almost all zero. One possible assumption to obtain a spin-dependent generation/recombination mechanism based on this selection rule is to assume that the exchange energy changes during the generation/recombination events vanishing at some point in time, for example for separated triplets for which the matrix elements  $|\langle n|S\rangle|$ become non-zero \cite{Collins2019}. It was also proposed that stochastic fluctuations of the exchange energy could also induce resonances enabling intersystem crossing \cite{collins2022quintet}. Here we decided to consider a simpler model, based on the mixing between quintet and singlet states due to the zero-field fine structure tensor of the triplet excitons. 
Treating this mixing in second order perturbation theory and assuming a large energy gap between the quintet and singlet manifolds, we arrive at the singlet mixing recombination model below.

{\it Model (II) } :
\begin{align}
  \alpha_n  &= \sum_{i} c_{i} | \; \langle S | \sum_\nu (\mathbf{u}_i \cdot \mathbf{{\hat S}^\nu})^2 | n \rangle \; |^2  \label{eq:phSS}
\\
  \Gamma_n  &=  \sum_{i} q_{i}  | \; \langle S | \sum_\nu (\mathbf{u}_i \cdot \mathbf{{\hat S}^\nu})^2 | n \rangle \; |^2
\label{eq:recSS}
\end{align}
where the sum runs over $i \in x,y,z$, the index $\nu = 1, 2$ corresponds to the two triplets forming the quintet state and $\mathbf{{\hat S}}^\nu$ to their spin operator. Indeed the matrix elements would still vanish if we used the total spin of the triplet pair in {\it model (II)} in the same way as we did in {\it model (I)}. Since we are looking at a $\pi$-stacked quintet we assume the rate constants $c_i, q_i$ to be independent of the triplet number index $\nu$. While this model was postulated based on perturbation theory arguments, the matrix elements appearing in Eqs.~(\ref{eq:phSS},\ref{eq:recSS}) will also appear inside rate models built to describe the effects of fluctuating exchange interactions. To summarize we introduced two spin-dependent generation/recombination models to attempt to model quantitatively the amplitude of the ODMR signal. The first model Eqs.~(\ref{eq:recS2gen},\ref{eq:recS2}) is inspired by the models for spin-orbit mediated phosphorescence of triplet excitons, while the second model takes into account the mixing between singlet and quintet states induced by the triplet fine structure parameters. The two models have the same number of adjustable parameters and it is thus reasonable to compare how they are able to reproduce the ODMR transition amplitudes.

In additional to {\it model (I)} and  {\it model (II)}, it is possible to consider mixed models where, for example, singlet mixing {\it model (II)} would be responsible for generation while intersystem crossing inspired {\it model (I)} would be responsible for recombination. However such models then have a very strong asymmetry between generation and recombination which will a-priori lead to many transitions of different sign which is not the case in the experiment where only one transition 23 becomes negative. While adjustment of rate parameters can reduce the number of sign changes, the fitting accuracy was in general lower and since we try to find only the dominant terms of the relaxation rates we will thus not discuss the results of mixed models in more detail. 

\section{Comparison with experiment}

\begin{figure}
\centerline{
  %
  %
  \includegraphics[clip=true,width=8cm]{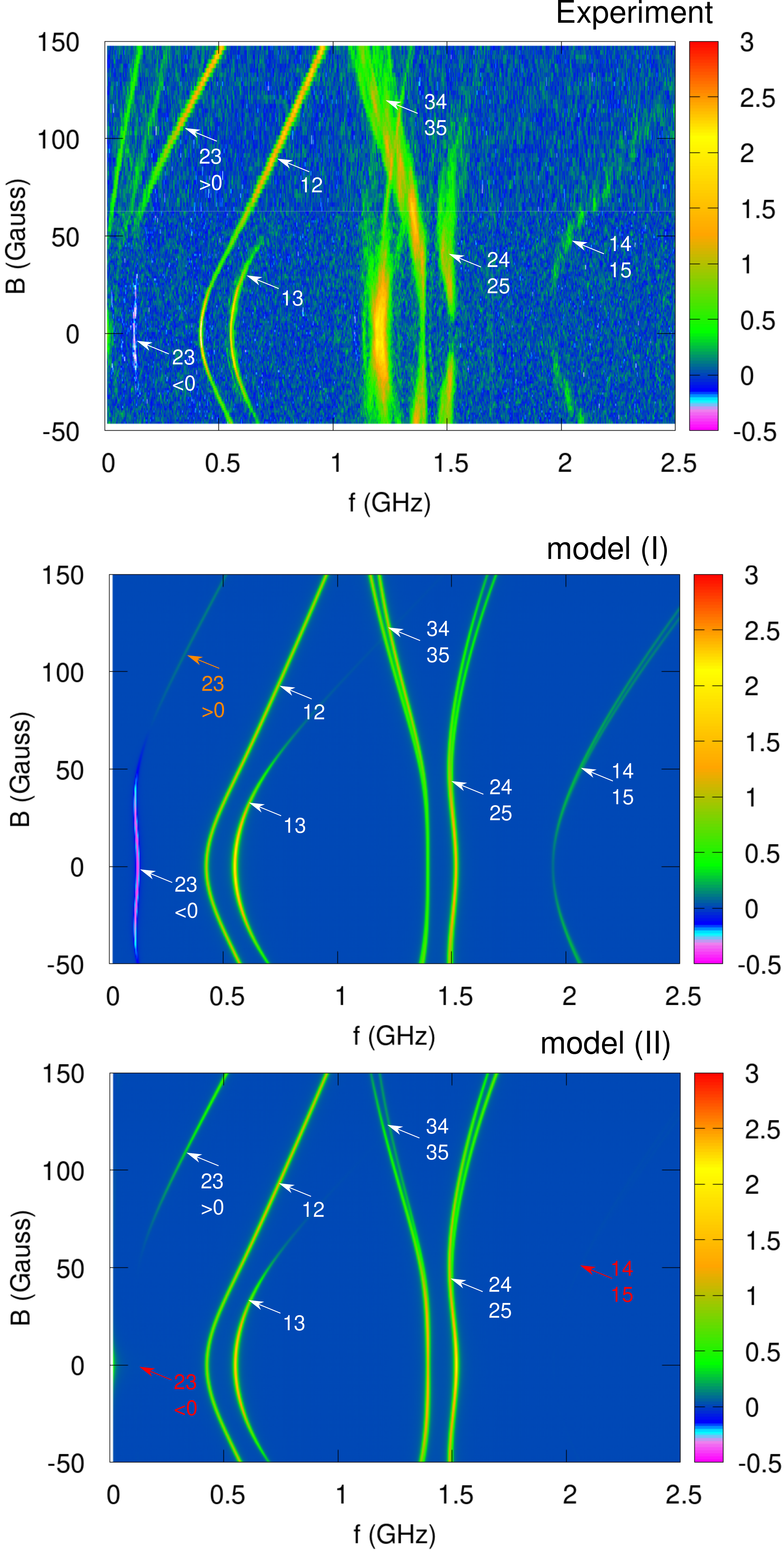}
}
\caption{Comparison between experimental broadband ODMR spectrum and numerical simulations using best fitting rates obtained for both {\it model (I)} and {\it model (II)}. The numerical rate values are summarized in Table \ref{TableSO}. Both models give very good general agreement with the experiment, however {\it model (II)} fails to predict the sign reversal of transition 23 and gives no visible amplitude for two spin flip transitions 14 and 15. 
}
\label{BestFitSO}
\end{figure}

In the previous section we described two possible kinetic models to simulate the amplitude of the ODMR signal from the underlying spin Hamiltonian of the system and we can now test if they can reproduce our 2D ODMR spectrum by adjusting the rate parameters. In order to find the best fitting rates in a systematic way, we extracted the ODMR amplitude for all the visible transitions, using a Lorentzian fit to the fixed magnetic field slices in Fig.~\ref{odmrExp}. We then used the rate equations to compute the peak amplitudes for each transition for varying magnetic field. The difference between rate equations and experiment was then estimate as a mean absolute deviation for all the ODMR amplitudes extracted from the experiment. This approach avoids having to compute the ODMR signal for every point of the two dimensional frequency/magnetic field maps reducing substantially the computational load of our optimization algorithm searching for optimal rates. This algorithm minimizes the mean ODMR amplitude error adjusting the kinetic rate parameters $a_i, r_i$ for {\it model (I)}  and the parameters $c_i, q_i$ for the singlet overlap {\it model (II)} as well as the spin independent rates $\alpha, \Gamma$.  For efficient minimization we used a controlled random search algorithm \cite{price1983global} provided by the NLopt package \cite{NLopt} which was found to have good convergence for this problem. The code for this optimization problem as well as for calculation of the triplet-pair Hamiltonian eigenvalues-eigenvectors is available at \cite{espin} We notice that so far the computed ODMR amplitudes depend only on the ratios between rates and not on their absolute value, to fix the timescale of spin-dependent recombination coefficients $\Gamma_n$ we will later use the dependence of the ODMR signal on the microwave modulation frequency. To compare the two proposed rate models, we do not yet need the absolute values of the rate coefficients and we will thus use a normalization where both the maximum spin-dependent generation and recombination are set to one, $\max_n \alpha_n = \max_n \Gamma_n = 1$. This normalization, (which adds an additional overall signal amplitude fitting parameter), allows direct comparison of the spin dependence of generation and recombination rates to see how much deviation from the relation $\alpha_n \propto \Gamma_n$ is needed to explain low field sign reversal for transition 23.

\begin{figure}[h]
\includegraphics[clip=true,width=8cm]{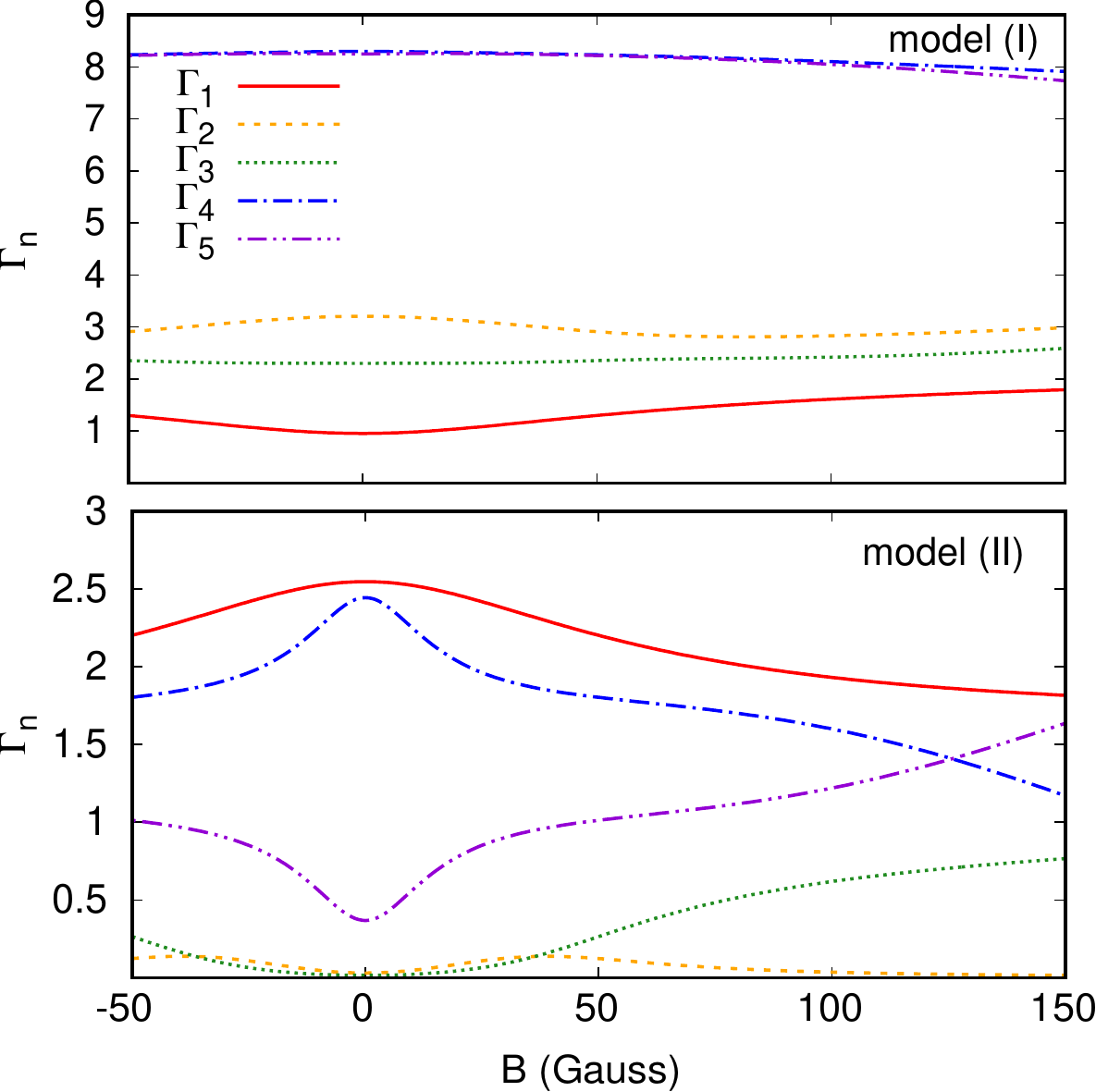}      
\caption{Best fitting relaxation rates $\Gamma_n$ for the two models, top figure shows the rates for {\it rate model (I)} using the rates Eqs.~(\ref{eq:recS2gen}, \ref{eq:recS2}), the bottom figure shows results for {\it rate model (I)} using Eqs.~(\ref{eq:phSS}, \ref{eq:recSS}). Since only ratios between rates are important for the ODMR signal amplitude the $y$ axis is in arbitrary units. To estimate the time length-scale associated with these rates we measured the microwave amplitude modulation frequency dependence of the ODMR signal for transitions 12 and 13 at zero magnetic field, these experiments suggest the order of magnitude $\Gamma_1 \simeq 60\;{\rm ms}^{-1}$.
}
\label{BestFitRates}
\end{figure}

\begin{center}
\begin{tabular}{ |c|c|c|c| } 
  \hline
  \multicolumn{4}{|c|}{ Model (I) } \\ \hline \hline
$\Gamma$ & $r_z$ & $r_y$ & $r_x$ \\ \hline
0 & 1 & 0.15 & 0 \\ \hline\hline
$ $ $\alpha$ $ $ & $ $ $a_z$ $ $ & $ $ $a_y$ $ $ &$ $ $a_x$ $ $\\ \hline 
0 & 1 & 0.2 & 0  \\  \hline 
 \hline
\end{tabular}
\quad
\begin{tabular}{ |c|c|c|c| } 
  \hline
  \multicolumn{4}{|c|}{ Model (II) } \\ \hline \hline
$\Gamma$ & $q_z$ & $q_y$ & $q_x$ \\ \hline
0.04 & 0.32 & 0.28 & 1 \\ \hline\hline  
 $ $ $\alpha$ $ $ & $ $ $c_z$ $ $ & $ $ $c_y$ $ $ &$ $ $c_x$ $ $\\ \hline
0.1 & 0.72 & 0 & 1  \\  \hline 
\hline
\end{tabular}
\end{center}
\captionof{table}{Summary of the optimal values of the spin-dependent recombination rates estimated for the intersystem crossing recombination {\it model (I)} and for the singlet-mixing model {\it model (II)}. Since the rate model depends mainly on ratios between rates, rates are scaled in a way that largest rate is set to unity. An estimation of the order of magnitude of rate constants is given in Fig.~\ref{BestFitRates}. As discussed in the main text the choice where $r_x = a_x = 0$ in {\it model (I)} corresponds to vanishing spin independent recombination/generation rates rather than large in plane anisotropy which is parameterized by $r_y, a_y > 0$. The rates are given in (relative) dimensionless units, dependence on the microwave square-wave modulation frequency shows that the recombination lifetimes are in the $100\;{\rm \mu s}$ range.
}\label{TableSO}.  
\begin{center}
\end{center}

Figure ~\ref{BestFitSO} shows the best fitting result obtained for {\it model (I)}. For the comparison between experiment and theory the magnetic fields were limited to $B \le 150\;{\rm Gauss}$, indeed in our rate models we assumed magnetic field independent relaxation rates. While this assumption seems plausible at weak magnetic fields it is likely not longer accurate when Zeeman energy becomes comparable with the zero-field energy splitting parameters $g \mu_B B \sim D$.

The best fitting rate parameters are given in Table \ref{TableSO}. One other fitting parameter not shown in the table is the angle between the strip line and the crystal $94^\circ$, which is consistent with the precision of the alignment of TIPS-tetracene crystal with the strip-line direction.
We see that overall the agreement between the experimental and simulated data for {\it model (I)} is very good. This model reproduces very accurately the amplitudes for transitions 12 and 23, it correctly captures the reversal of the signal amplitude at low magnetic fields for transition 23, and the weak but visible signal for the transitions 14, 15. The main disagreement between experiment and simulations seems to be the amplitude of transitions 24, 25 whose amplitude vanishes at zero magnetic field. We see in Fig.~\ref{odmrLevels} that the levels 4 and 5 are almost degenerate at zero magnetic field and it is possible that hyperfine effects need to be included in the spin Hamiltonian to explain this discrepancy.

The algorithm finding the best fitting rate values for {\it model (I)} converged to several vanishing terms $r_x = a_x = \Gamma = \alpha = 0$ and $r_y \simeq a_y$ (we note that for now $r_z = a_z = 1$ by normalization), with only a small difference between non-zero in generation/recombination rates $a_y = 0.2$, $r_y = 0.15$ which is sufficient to explain the zero-field sign reversal for transition 23. Since $a_y > 0$, we need to comment on $a_x = 0$ as these values seem highly anisotropic. For {\it model (I)}, we note the identity ${\hat S}_x^2 + {\hat S}_y^2 + {\hat S}_z^2 = S(S+1)$ where $S = 2$ is total spin, thus for $a_z > a_y > a_x$ a finite $a_x > 0$ would actually describe spin-independent recombination which is also contained in another term $\alpha$ (the optimization also yields $\alpha=0$). 
The direction of the dominant relaxation rate $r_z$ coincides with the main direction of the triplet fine-structure tensor giving the $D_Q S_z^2$ term of the quintet spin-Hamiltonian while the difference between the rates $a_x, r_x$ and $a_y, r_y$ reflects the in plane anisotropy of the quintet. This value is comparable to the splitting $(\epsilon_3-\epsilon_2)/D_Q \simeq 0.25$ induced by the dipole-dipole interaction between triplets in the quintet spectrum, which gives the finite $E_Q$ value. We notice that we find $a_y > r_y$, the larger value of $a_y$ can be a sign of transient deformations during the fast generation through the fission process to which the slower recombination processes are not sensitive. 

As discussed above the basis retained for this model was the zero-field quintet basis which does not exactly coincide with the molecular basis of the TIPS-tetracene molecules. This basis choice was motivated by the strong exchange energy for the $\pi$ stacked quintet which projects all the matrix elements into the quintet manifold. It is also possible to write the same kinetic model in the molecular basis of one of the triplets (see Fig.~\ref{odmrSketch}), this change of basis will lead to a change of all the matrix elements and thus of best fitting kinetic rates. Running the optimization algorithm for this choice leads to a similar theoretical 2D ODMR spectrum but with more non-vanishing rates for the generation  $a_{x,y,z}, \alpha$ and recombination rates $r_{x,y,z}, \Gamma$. These additional non-zero rates in the molecular basis probably come from the rotation between the two frames, as a confirmation of this we checked that predicted spin populations were also equal for these two version of {\it model (I)} confirming that these are two pictures of the same physics in two different frames. We thus feel that these results justify use of the quintet zero field basis in {\it model (I)}. 

For the singlet-projection {\it model (II)}, the best fitting results are also displayed on Figure ~\ref{BestFitSO} with the corresponding rate parameters given in Table \ref{TableSO}. We see that this model leads to amplitudes which are very close to the values obtained from the spin-orbit interaction model for transitions 12, 23, 24, 25, 34, 35. However, in disagreement with the experiment, this model predicts vanishing signal amplitudes for the transitions 23, 14, 15. Also most of the estimated rates are non zero with dominant rates $c_x = q_x = 1$ with a surprisingly strong anisotropy between $x$ and $y$ rates, which is hard to justify and very different values between generation and recombination (normalized) rates. The simplicity of the results of {\it model (I)} and its better agreement with experiment suggest that this model may indeed provide, the dominant spin-dependent recombination pathway. It is however also instructive to understand physically why {\it model (II)} is unable to account for the amplitudes of some of the transitions.

To understand the difference between the two models it is helpful to consider the dependence of the spin-dependent recombination rates for all the states of the quintet as a function of the magnetic field, it is displayed on Fig.~\ref{BestFitRates}. We see that for {\it model (II)} both rates $\Gamma_2$ and $\Gamma_3$ tend to vanish near zero magnetic field giving vanishing ODMR contrast for the transition 23 near zero magnetic field, which is not the case for {\it model (I)} where there is a sign reversal.
These vanishing rates are actually implied by an unexpected selection rule for {\it model (II)}:
\begin{align}
\langle S| ({\hat S}_{x,y,z}^{\nu})^2 | \pm 1_Q \rangle = 0
\end{align}  
which holds because $|0\rangle_T$ is always an eigenvector of the triplet spin square operators with either $({\hat S}_{x,y}^{\nu})^2 |0\rangle_T = |0\rangle_T$ and $({\hat S}_{z}^{\nu})^2 |0\rangle_T = 0 |0\rangle_T$ (with $\nu = a,b$ the two triplets of the biexciton pair).

Another difference between the two models is the larger ODMR amplitude for transitions 14 and 15 that is obtained for {\it model (I)} compared to  {\it model (II)} in better keeping with experiment. The two transitions 14 and 15 correspond to the $|0\rangle_Q \rightarrow |\pm 2\rangle_Q$ transitions with a change of total spin $z$-projection by two quanta of momenta. Such transitions are only allowed because of the weak in plane anisotropy and thus have much smaller amplitudes compared to the other transitions. To observe them, a high ODMR contrast between the photoluminescence rates for $\Gamma_{4,5}$ and $\Gamma_1$ is needed to compensate for the weak transition amplitudes. As seen in Fig.~\ref{BestFitRates}, such contrast is naturally provided by the $\langle n|{\hat S}_z^2|n \rangle$ matrix elements for {\it model (I)} with much larger expectation values for $\pm 2$ states compared to other states. The fact that transitions 14 and 15 are observable in the experiment with amplitudes not much weaker than the other selection-rule-allowed transition is thus a strong argument in favor of {\it model (I)}.

Note that in Fig.~\ref{BestFitRates}, we give the rates in arbitrary units. The time-scale for spin-dependent recombination rates in Fig.~\ref{BestFitRates} can be estimated from the dependence of the ODMR signal with microwave square modulation frequency. We fixed the rate scale using the quintet transitions 12 and 13 at zero magnetic field. This suggest that spin-dependent fluorescence recombination life-times are in the $100\;{\rm \mu s}$ range, which is indeed a long timescale for pulsed magnetic resonance experiments. 

\begin{figure}
\includegraphics[clip=true,width=8cm]{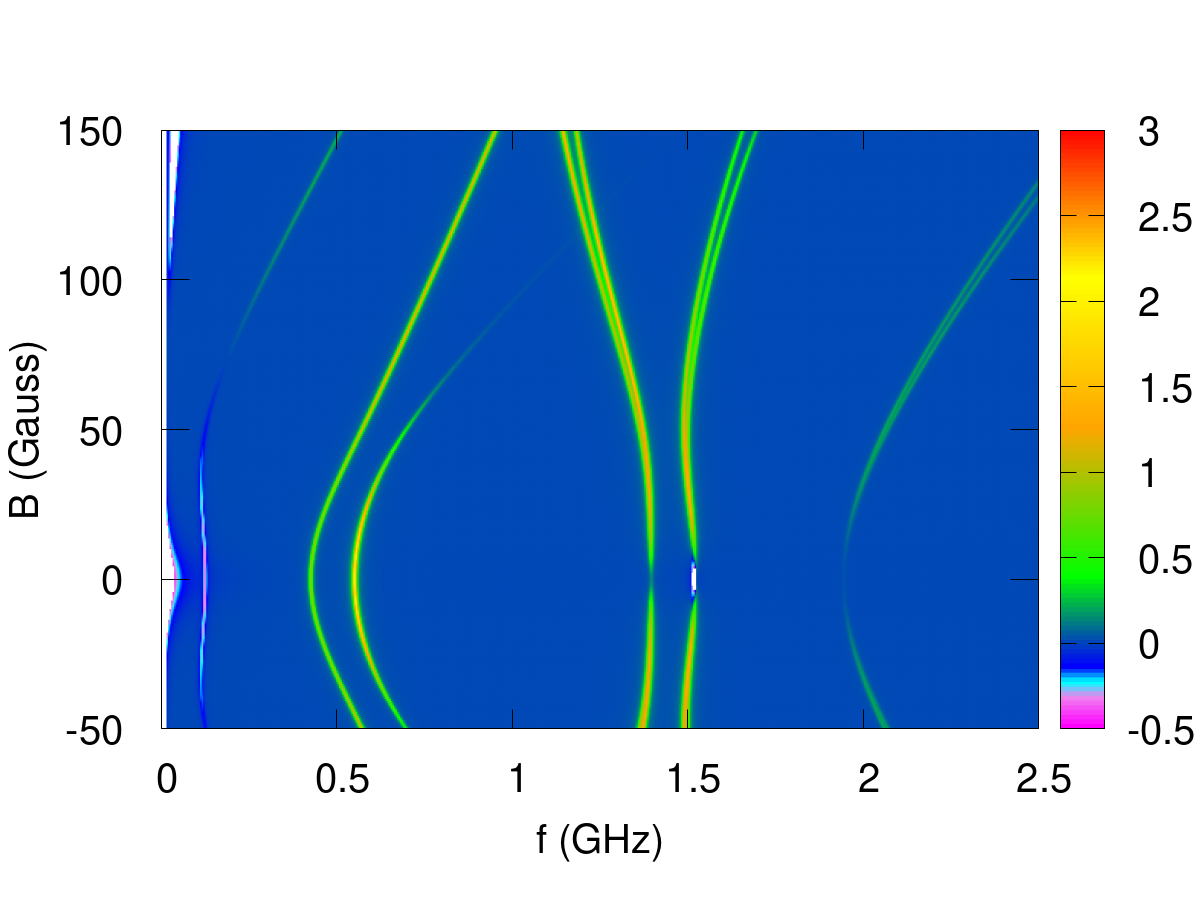}  
\caption{Simulated broadband ODMR spectrum for the zero field basis relaxation {\it model (III)}. This model captures the zero field $B = 0$ anomaly at 1.5 GHz which is also observed in the experiment. However, this is achieved at the cost of introducing more rate parameters which are all non zero after optimization compared to {\it model (I)}, which this model generalizes. Population and generation/recombination rates of the spin eigenstates as function of magnetic field are similar to those obtained for {\it model(I)} (shown on Fig \ref{BestFitRates})}
\label{SimuZFH}
\end{figure}

To conclude the discussion on the possible models for ODMR amplitudes and comparison with experiment, we notice that the agreement with {\it model (I)} is not perfect with differences concerning transitions 24 and 25 which exhibit a dip in amplitude very close to zero field. These transitions also show a faster decay with magnetic field but this can have extrinsic explanations like a higher sensitivity to sample inhomogeneity. On the contrary the zero field behavior must be intrinsic as the effect of inhomogeneous broadening is minimal at zero magnetic field and we thus concentrate on the zero field region. Since the magnetic field scales involved are very small, $\sim 5$ Gauss, this probably implies that the relaxation mechanism is sensitive to the weak mixing between $|\pm 2\rangle$ states which is induced by quintet $E_Q$ in plane anisotropy. This mixing does not induce any changes in the matrix elements for both model {\it model (I)} nor {\it model (II)} and thus this feature is missing from simulations in Fig.~\ref{BestFitSO}. A way to increase the sensitivity to this mixing is to generalize {\it model (I)} by assuming that the relaxation kinetics are not only sensitive to the expectation values of some spin operators in the quintet zero field basis but directly to the zero field quintet eigenvectors.


This leads to {\it Model (III) } :
\begin{align}
  \alpha_n  &= \sum_{m} d_{m} | \langle m(B=0) | n \rangle |^2  \label{eq:phB0}
\\
  \Gamma_n  &=  \sum_{m} u_{m}  | \langle m(B=0) | n \rangle |^2 \label{eq:recB0}
\end{align}
where the sums runs over the five zero quintet zero field eigenvectors $| m(B=0) \rangle$ and ${d_m, u_m}$ a set of 10 spin-dependent generation/recombination rates. This model can be viewed as a generalization of {\it model (I)} under the principle that all rates are determined by zero field eigenstates. It has more fitting parameters, and it is thus expected to preform at least as well. Contrarily to the matrix elements of {\it model (I)}, we don't see any heuristic microscopic justification for this generalization, and it looses the natural explanation for the strong contrast between $|\pm 2\rangle$ and other eigenstates. This contrast has to be obtained by adjusting the rates of the eigenstates. Losing these advantages of {\it model (I)} to explaining a zero field anomaly in transitions 24, 25 may seem excessive. Thus we decided not focus on {\it model (III)}, we just show in Fig.~\ref{SimuZFH} that it can indeed reproduce a zero field anomaly for transitions 24 and 25 with a magnetic field scale comparable to the experiments. The presence of some evidence that in some cases we may need to generalize {\it model (I)} to {\it model (III)} seems worth noting.

\begin{figure}
\includegraphics[clip=true,width=8cm]{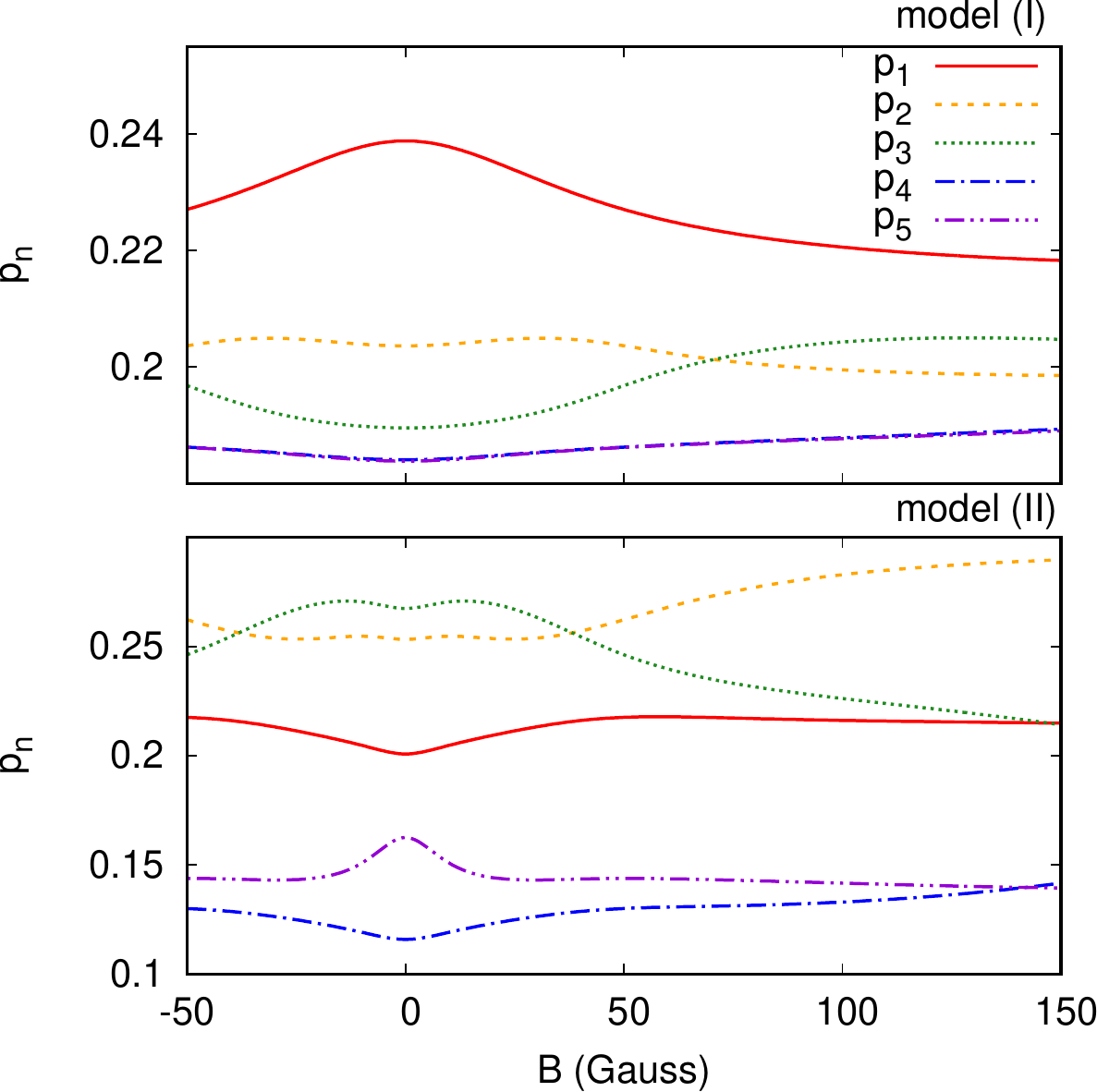}  
\caption{Simulated normalized populations $p_n = P_n / \sum_{j} P_j$ for {\it model (I)} and {\it model (II)}, {\it model (I)} does not predict any population inversion at zero magnetic field for this system, while population inversion would be expected for {\it model (II)}. The populations and rates predicted for  {\it model (I)} were robust with respect to rotation of the molecular basis in which the rate model was written and changed only weakly after upgrading to the more general {\it model (III)}. The results for {\it model (II)} where less robust with respect to basis rotation (see main text).}
\label{SimuPop}
\end{figure}

Using the rate values from \ref{TableSO}, it is possible to reconstruct the expected population of the quintet spin sub-levels as function of magnetic field for {\it kinetic models (I)-(II)} using Eqs.~(\ref{eq:recS2gen}, \ref{eq:recS2}) and rate equations. We show the corresponding results in Fig.~\ref{SimuPop} as function of the magnetic field. To avoid dependence on the overall generation rates the populations are given as occupation probabilities $p_n = P_n / \sum_{j} P_j$.

For model {\it model (I)}, we find that there is no population inversion around zero magnetic fields. At higher magnetic fields the populations of states 2 and 3 do invert. This inversion in population corresponds to the ODMR signal switching from negative to positive. Note that the possibility of finding population inversion regimes without relying on microwave resonant circuits could be useful to find new gain materials for organic MASERs \cite{oxborrow2012room,bogatko2016molecular,wu2022enhanced}. As we mentioned previously we found this population to be robust to changes of the preferential directions in which {\it model (I)} was written switching from quintet-zero field to the molecular frame. Except for the relative populations of states 4,5 the populations were also unchanged in {\it model (III)}. This strengthens our conclusions that {\it model (I)} is probably the dominant spin-dependent radiative recombination pathway for quintet bi-excitons. For completeness, we also show the predicted population for {\it model (II)}, the predicted occupation probability for this model is very different, but not very stable with respect to small variations in {\it model (II)}. Thus based on this analysis, the population and relaxation rates from {\it model (I)} seem the most likely. Conceptually it is possible to try to confirm the predictions of this model experimentally, by performing a three-dimensional ODMR spectrum where the microwave amplitude modulation (AM) frequency dependence of the signal would also be recorded for each transition while both magnetic field and microwave frequencies are changed. This would be certainly very demanding both in terms of experiment and data analysis, so we used only the zero field AM frequency dependence to fix the order of magnitude of the rates in Fig.~\ref{BestFitRates}. Indeed our aim here, was to make as much progress as possible based on the analysis of the two dimensional ODMR spectrum revealing the rich physics that is hidden and that was not explored previously. 

\section{Conclusions}

We investigated the mechanisms for spin-dependent photoluminescence for quintet bi-excitons formed through singlet fission, by studying the broadband optically detected magnetic resonance (ODMR) signal from a quintet bi-exciton formed by two triplet excitons on nearby $\pi$-stacked TIPS-tetracene molecules. Our broadband excitation and readout schemes allowed us to measure  the ODMR signal for all the 10 possible quintet transitions as a function of magnetic field. We then showed that competing spin-dependent pathways are needed to explain the sign change of the ODMR amplitude, which is observed for one of the transitions. We then proposed two possible spin-dependent recombination models. The first model was based on the matrix elements of the fine structure quintet operators which generalizes the zero field basis intersystem crossing models which were introduced for triplet excitons. In the second model we considered the mixing of the quintet states with the singlet wavefunction  which is induced perturbatively by the triplet fine-structure terms. 
While both models were successful in reproducing the ODMR amplitude of most of the transitions, we found that only the first model made accurate predictions for two transitions in the manifold. Spin forbidden transitions which were observed despite their weak magnetic resonance cross-section require a strong optical contrast to make them visible. This large contrast was naturally obtained in the first model while it is missing in the singlet-mixing model. The second model was further unable to explain the sign reversal of one of the transitions near zero magnetic field as a hidden selection rule in this model imposes vanishing optical contrast for this transition near zero magnetic field. While the first model succeeded in reproducing the experimental data with a minimal number of parameters with only a weak difference between generation and recombination rates. Thus our experiments and analysis provide a strong indication that the matrix elements of the quintet fine-structure term describes the dominant spin-dependent generation/recombination pathway for quintet bi-excitons. 

Acknowledgments: This project was supported by funding from ANR-20-CE92-0041 (MARS). L.R. Weiss acknowledges support from the University of Chicago/Advanced Institute for Materials Research Joint Research Center. V. Derkach acknowledges support from PHC DNIPRO (Project 46809UD) and Poncelet center, as well as the kind hospitality from CNRS Gif-sur-Yvette. 

\bibliography{ampodmr.bib}

\end{document}